%% file: CMW_submit_2.tex
\documentclass[twocolumn,showpacs,prl,superscriptaddress]{revtex4-1}
\def\pt{\mbox{$p_T$}}   
\def\sNN{\mbox{$\sqrt{s_{_{\rm NN}}}$}}   
\newcommand{ \be }{\begin{equation}}       
\newcommand{ \ee }{\end{equation}}       
\newcommand{ \bea }{\begin{eqnarray}}       
\newcommand{ \eea }{\end{eqnarray}}

\newcommand{ \etal }{{\it et al.}}   
\newcommand{\ach}{A_{\textrm{ch}}}
\usepackage{color}

\usepackage{graphicx} 
\usepackage{fixmath} 
\usepackage{dcolumn} 
\usepackage{bm} 
\usepackage{multirow}
\usepackage{ulem}
\usepackage[squaren]{SIunits}

\usepackage[absolute]{textpos}
\textblockorigin{20mm}{10mm} 
\begin{document}          


\title{Observation of charge asymmetry dependence of pion elliptic flow and the possible chiral
  magnetic wave in heavy-ion collisions}
 
\include{star_authors}

\begin{abstract}     
  We present measurements of $\pi^-$ and $\pi^+$ elliptic flow, $v_2$, at
  midrapidity in Au+Au collisions at $\sNN =$ 200, 62.4, 39, 27, 19.6, 11.5 and 7.7~GeV, as a
  function of event-by-event charge asymmetry, $\ach$, based on data from the STAR experiment
  at RHIC. We find that $\pi^-$ ($\pi^+$) elliptic flow linearly increases (decreases) with
  charge asymmetry for most centrality bins at $\sNN = \text{27~GeV}$ and
  higher. At $\sNN = \text{200~GeV}$, the slope of the difference of $v_2$ between $\pi^-$ and $\pi^+$
  as a function of $\ach$ exhibits a centrality dependence, which is qualitatively similar to
  calculations that incorporate a chiral magnetic wave effect. Similar centrality dependence is
  also observed at lower energies.
\end{abstract} 
 
\maketitle  

In heavy-ion collisions at the Relativistic Heavy Ion Collider (RHIC) and the Large Hadron
Collider (LHC), energetic spectator protons produce a strong magnetic field reaching
$eB_y \approx m^2_\pi$~\cite{Kharzeev}, or $\sim 3 \times 10^{14}$~T. The interplay between the
magnetic field and the quark-gluon matter created in these collisions might result in two
phenomena: the chiral magnetic effect (CME) and the chiral separation effect (CSE).  The CME is
the phenomenon of electric charge separation along the axis of the magnetic field in the
presence of a finite axial chemical
potential~\cite{Kharzeev,Kharzeev2,Kharzeev3,Kharzeev4,Kharzeev5}. The
STAR~\cite{STAR_LPV1,STAR_LPV2,STAR_LPV3,STAR_LPV4} and PHENIX~\cite{PHENIX_LPV1,PHENIX_LPV2}
Collaborations at the RHIC and the ALICE Collaboration at the LHC~\cite{ALICE_LPV} have reported
experimental observations of charge separation fluctuations, possibly providing evidence for the
CME. This interpretation is still under discussion (see e.g.~\cite{dis1,dis2,Adamczyk2014m} and
references therein). The CSE refers to the separation of chiral charge, which characterizes
left/right handedness, along the axis of the magnetic field in the presence of the finite
density of electric charge~\cite{CSE1,CSE2}. In this Letter, we report the results from a search
for these effects using a new approach.

In a chirally symmetric phase, the CME and CSE can form a collective excitation, the chiral
magnetic wave (CMW). It is a propagation of chiral charge density in a long wavelength
hydrodynamic mode~\cite{CMW,CMW2,Gorbar2011a}. The CMW, which requires chiral symmetry restoration,
manifests itself in a finite electric quadrupole moment of the collision system, where the
``poles" (``equator") of the collision system acquire additional positive (negative)
charge~\cite{CMW}.  This effect, if present, will increase(decrease) the elliptic flow of
negative (positive) particles. Elliptic flow refers to an azimuthally anisotropic collective
motion of soft (low momentum) particles. It is characterized by a second-order harmonic in a
particle's azimuthal distribution, $\phi$, with respect to the reaction plane azimuthal angle,
$\Psi_{\rm RP}$, which is determined by the impact parameter and the beam direction,
\begin{equation}{}
  v_2 = \langle \cos[2(\phi - \Psi_{\rm RP})] \rangle.
\end{equation}
The CMW is theoretically expected to modify the elliptic
flow of charged particles, e.g. pions, on top of the baseline
$v_2^{\rm base}(\pi^\pm)$~\cite{CMW}
\begin{equation}
  v_2(\pi^\pm) = v_2^{\rm base}(\pi^\pm) \mp \frac{r}{2}\ach,
\end{equation}
where $r$ is the quadrupole moment normalized by the net charge density and
$\ach = (N_+ - N_-)/(N_+ + N_-)$ is the charge asymmetry of the collision system. As the
colliding nuclei are positively charged, the average charge asymmetry, $\langle \ach \rangle$,
is always positive. Thus, the $\ach$-integrated $v_2$ of $\pi^-$ ($\pi^+$) should be above
(below) the baseline because of the CMW.  However, the $v_2^{\rm base}$ may be different between
$\pi^+$ and $\pi^-$ because of several other possible physical
mechanisms~\cite{quarkTransport,hadronicPotential,hybrid,partonicMeanField}. It is
preferable to study CMW via the $\ach$ dependence of the pion $v_2$ other than
$\ach$-integrated $v_2$.

This Letter reports the $\ach$-differential measurements of the pion $v_2$, based on Au+Au
samples of $2\times10^8$ events at 200 GeV from RHIC year 2010, $6 \times 10^7$ at 62.4 GeV
(2010), $10^8$ at 39 GeV (2010), $4.6\times10^7$ at 27 GeV (2011), $2\times10^7$ at 19.6 GeV
(2011), $1\times 10^7$ for 11.5 (2010) and $4\times 10^6$ for 7.7 GeV (2010). All events were
obtained with a minimum-bias trigger which selects all particle-producing collisions, regardless
of the extent of overlap of the incident nuclei~\cite{Bieser2003766}. Charged particle tracks
with pseudorapidity $|\eta|<1$ were reconstructed in the STAR Time Projection Chamber
(TPC)~\cite{TPC-NIM}. The number of charged particles within $|\eta| < 0.5$ is used to define
the centrality. The centrality definitions and track quality cuts are the same as those used in
Ref.~\cite{BESv2}, unless otherwise specified. Only events within \unit{40}{cm} (\unit{50}{cm}
for \unit{11.5}{GeV} and \unit{70}{cm} for \unit{7.7}{GeV}) of the center of the detector center
along the beam line direction are selected. To suppress events from collisions with the beam
pipe (radius = 3.95 cm), a cut on the radial position of the reconstructed primary vertex within
2 cm was applied.  A cut on the distance of the closest approach to the primary vertex (DCA
$< 1$ cm) was applied to all tracks to suppress contributions from weak decays and/or secondary
interactions.

\begin{figure}[t]
  \includegraphics[width=0.50\textwidth]{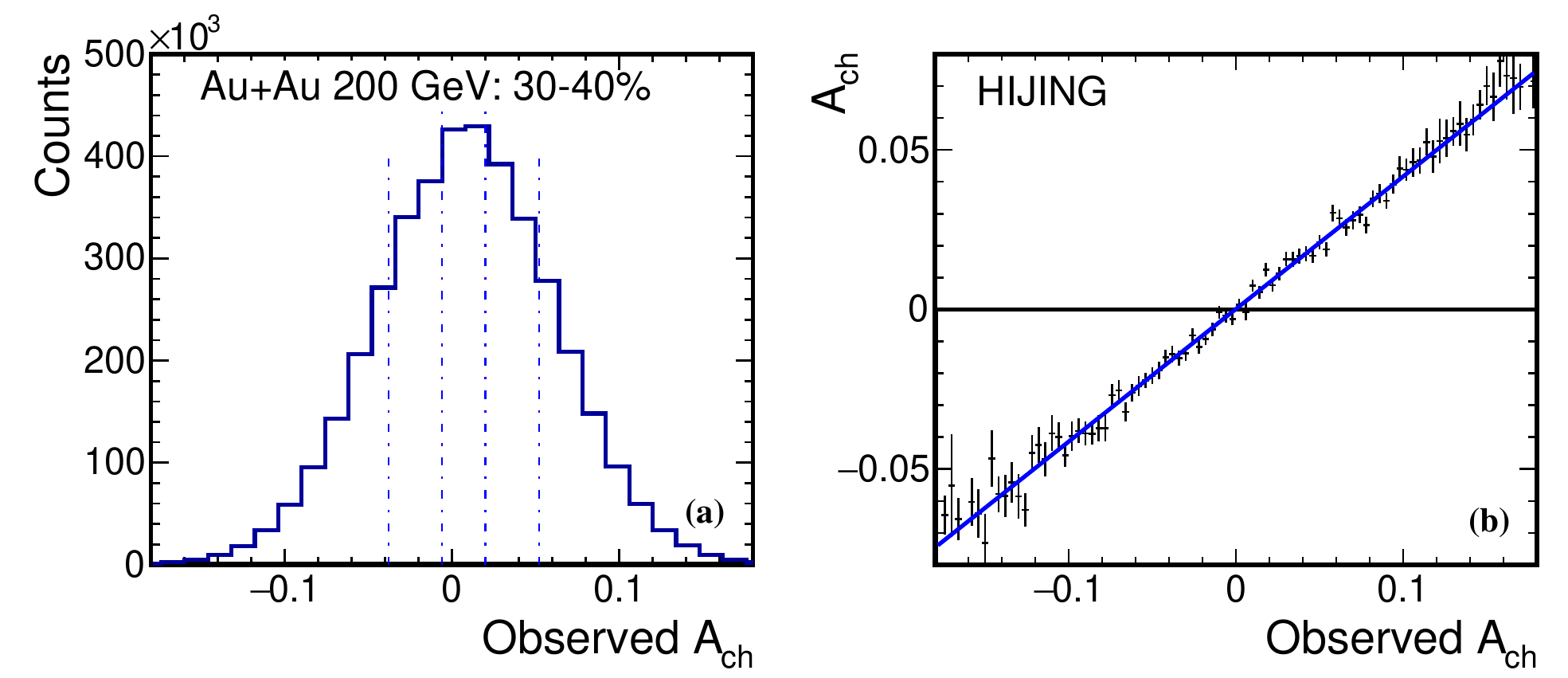}
  \caption{ (Color online) (a) the distribution of observed charge asymmetry from STAR data and,
    (b) the relationship between the observed charge asymmetry and the charge asymmetry from
    HIJING generated events, for $30$-$40\%$ central Au+Au collisions at 200
    GeV. In this centrality, the mean charge asymmetry $\langle \ach \rangle$
      of HIJING events is about 0.004. The errors are statistical only.}
    \label{fig:A}
\end{figure}

The observed $\ach$ was determined from the measured charged particles with transverse momentum
$p_T>0.15$ GeV/$c$ and $|\eta|<1$; protons and anti-protons with $p_T<0.4$ GeV/$c$ were excluded
to reject background protons from the nuclear interactions of pions with inner detector
materials. Fig.~\ref{fig:A}(a) shows an example of the observed $\ach$ distribution, which was
divided into five samples roughly containing equal numbers of events, as indicated by the dashed
lines.  Fig.~\ref{fig:A}(b) shows the relationship between the observed $\ach$ and the $\ach$
from the HIJING event generator~\cite{HIJING}, where the same cuts as used in data were applied
to calculate $\ach$. The relationship is linear.
To select pions with high purity, we eliminate charged particles more than
$2\sigma$ away from the expected energy loss of pions in the TPC. For energies less than or
equal to \unit{62.4}{GeV}, elliptic flow measurements were carried out with the $v_2$\{$\eta$
sub\} approach~\cite{Poskanzer1998}, where two subevent planes register charged particles with
$\eta>0.3$ and $\eta<-0.3$, respectively. Pions at positive (negative) $\eta$ are then
correlated with the subevent plane at negative (positive) $\eta$ to calculate $v_2$. The $\eta$
gap of 0.3 unit suppresses several short-range correlations such as the Bose-Einstein
interference and the Coulomb final-state interactions~\cite{Flow200GeV}.  There are correlations
that are unrelated to the reaction plane that are not suppressed by the $\eta$ gap, e.g. those
due to back-to-back jets. These are largely canceled in the $v_2$ difference between $\pi^-$ and
$\pi^+$. For \unit{200}{GeV}, the two-particle cumulant method
$v_2\{2\}$~\cite{Methods,Flow200GeV} was employed, which was consistent with $v_2$\{$\eta$
sub\}, and allowed the comparison with the $v_2\{4\}$ method discussed later in this letter. The
same $\eta$ gap was also used in the $v_2\{2\}$ analysis. To focus on the soft physics regime,
only pions with $0.15 < p_T < 0.5$ GeV/$c$ were used to calculate the $p_T$-integrated $v_2$,
and this $p_T$ range covers $65$-$70\%$ of all the produced pions.  The calculation of the
$p_T$-integrated $v_2$ was corrected with the $p_T$-dependent tracking efficiency for pions.

\begin{figure}[t]
  \includegraphics[width=0.50\textwidth]{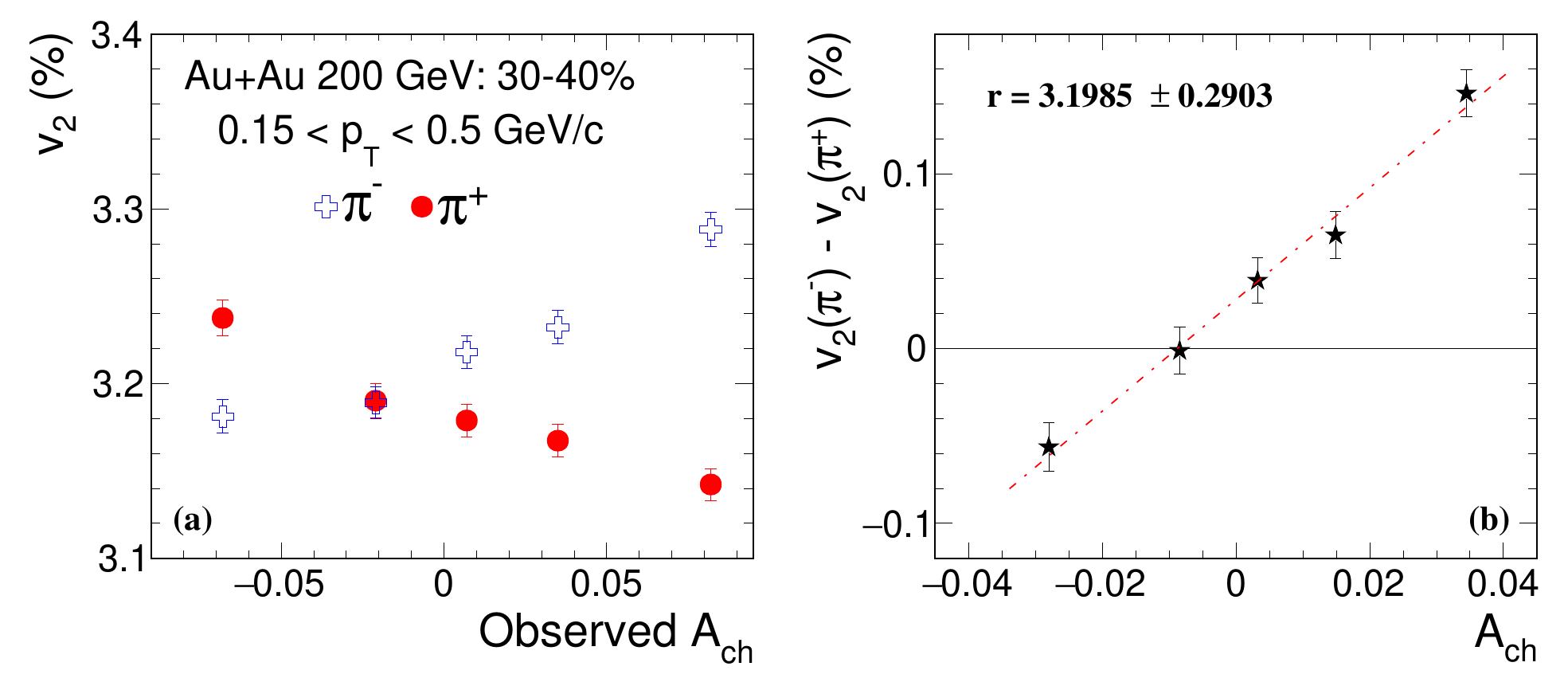}
  \caption{ (Color online) (a) pion $v_2\{2\}$ as a function of observed charge asymmetry and,
    (b) $v_2$ difference between $\pi^-$ and $\pi^+$ as a function of charge asymmetry with the
    tracking efficiency correction, for $30$-$40\%$ central Au+Au collisions at 200 GeV. The
    errors are statistical only.}
    \label{fig:example}
\end{figure}

\begin{figure}[t]
  \includegraphics[width=0.5\textwidth]{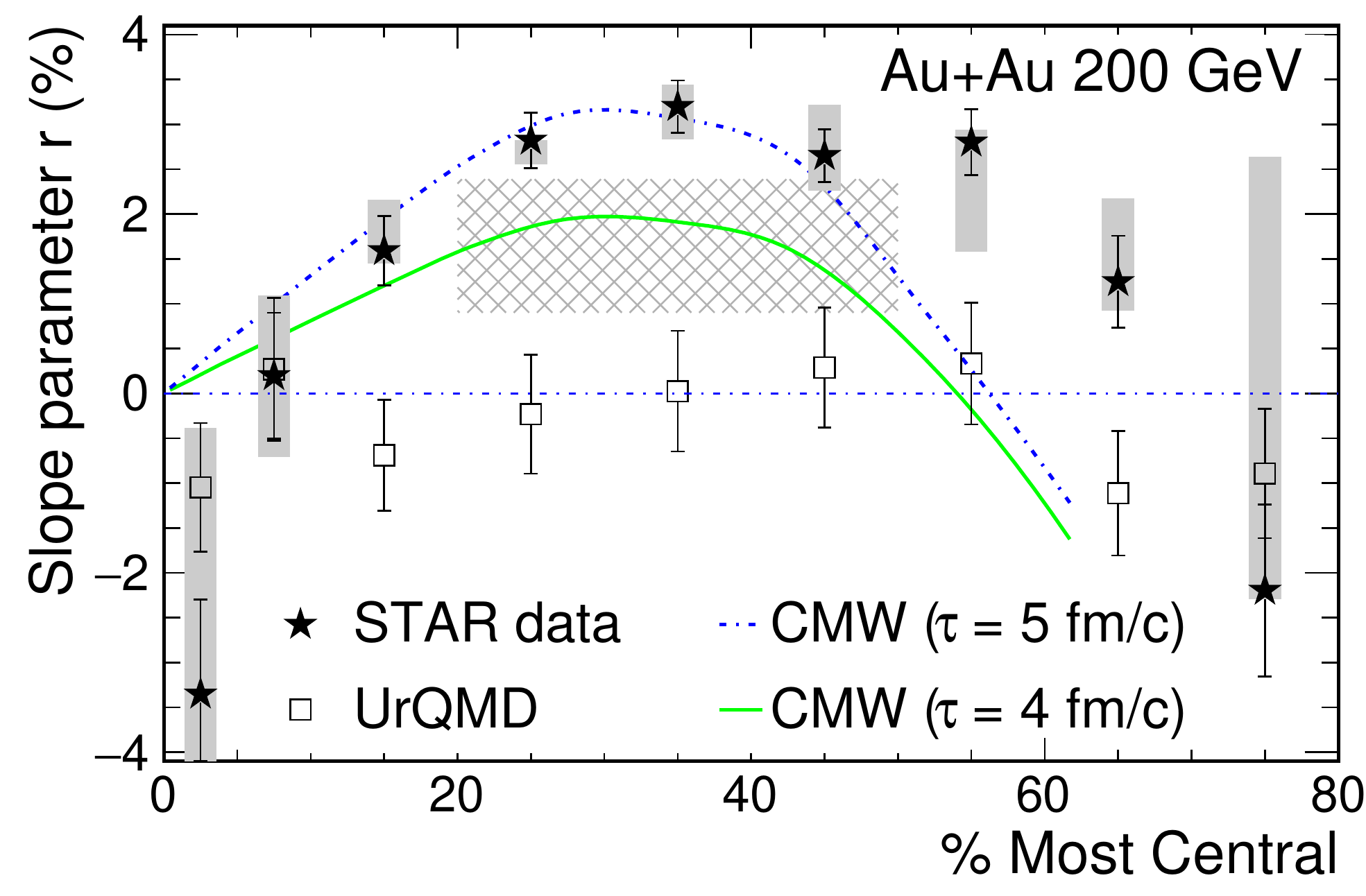}
  \caption{(Color online) The slope parameter, $r$, as a function of centrality for Au+Au
    collisions at 200 GeV. Also shown is the UrQMD~\cite{urqmd} simulation, and the calculations
    with CMW~\cite{CMWnewSlope} with different duration times.  The grey bands include the
    systematic errors due to the DCA cut, the tracking efficiency and the $p_T$ range of
    particles involved in the event plane determination. The cross-hatched
    band indicates the STAR measurement with the $v_2\{4\}$ method and the height of this band
    shows only the statistical error.} \label{fig:200GeV}
\end{figure}

Taking Au+Au 200 GeV collisions in the $30$-$40\%$ centrality range as an example, the pion $v_2$ is
shown as a function of the observed $\ach$ in Fig.~\ref{fig:example}(a). The $\pi^-$ $v_2$
increases with increasing observed $\ach$ while the $\pi^+$ $v_2$ decreases with a similar magnitude
of the slope. After applying the tracking efficiency to $\ach$, the $v_2$
difference between $\pi^-$ and $\pi^+$ has been fitted with a straight line as shown in
Fig.~\ref{fig:example}(b). The slope parameter, $r$, from Eq.~2, is positive and qualitatively
consistent with the expectations of the CMW picture. The fit function is
  non-zero at the average charge asymmetry $\langle \ach \rangle$, which is a small positive
  number in case of Au + Au collisions. This indicates the $\ach-$integrated $v_2$ for $\pi^-$
  and $\pi^+$ are different, which was observed in Ref.~\cite{BESv2_PID}.
We follow the same procedure as above to extract the slope parameter, $r$, for all centrality
bins at 200 GeV.  The results are shown in Fig.~\ref{fig:200GeV}, together with simulations
using the UrQMD event generator~\cite{urqmd} and with the theoretical calculations with
CMW~\cite{CMWnewSlope} with different duration times of the magnetic field.
For most data points, the slopes are positive and reach a maximum in mid-central/mid-peripheral 
collisions, a feature also seen in the theoretical calculations of the CMW. 
The gray bands in Fig.~\ref{fig:200GeV} include three types of systematic errors: the DCA cut
for pion tracks was tightened to 0.5 cm, to study the contribution from weak decays,
which dominates the systematic errors; the tracking efficiency for charged
particles was varied by relative $5\%$, to determine the uncertainty of $\ach$; and the $p_T$
range of particles involved in the event plane determination was shrunk from $[0.15, 2]$ GeV/$c$
to $[0.7, 2]$ GeV/$c$, to further suppress short-range correlations.  The $\ach$ bin center
may not accurately reflect the true center of each $\ach$ bin in Fig.~\ref{fig:example}, as
the $v_2$ measurements are effectively weighted by the number of particles of interest.  Such an
uncertainty on $r$ has been estimated to be negligible for most centrality bins, except for the
most peripheral collisions, where this systematic error is still
much smaller than the statistical error.
\begin{figure}[t]         
  \includegraphics[width=0.5\textwidth]{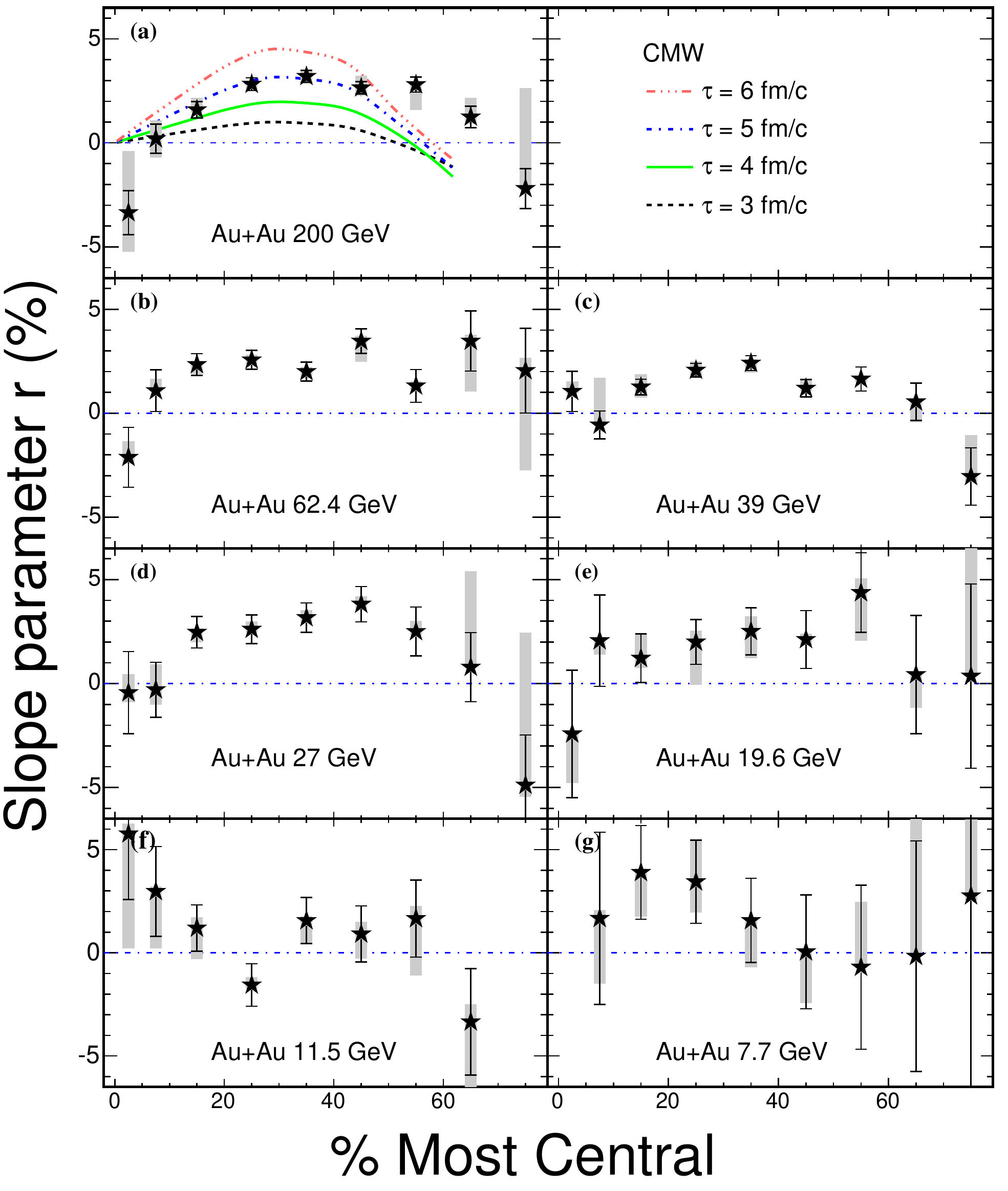}      
  \caption{(Color online) The slope parameter $r$ as a function of centrality for all the
    collision energies under study.  For comparison, we also show the calculations with
    CMW~\cite{CMWnewSlope} with different duration times.  The grey bands carry the same meaning
    as those in Fig.~\ref{fig:200GeV}.  } \label{fig:BES}
\end{figure}  

To further study the charge-dependent contribution from jets and/or resonance decays, we
separated positive and negative particles in each subevent to form positively (negatively)
charged subevents. Then each $\pi^+$ ($\pi^-$) is only correlated with the positive (negative)
subevent in the opposite hemisphere. The slope parameters thus obtained are statistically
consistent with the previous results though with larger uncertainties.

The event plane reconstructed with particles recorded in the TPC approximates the participant
plane, the measured $v_2$ are not the mean values, but closer to the root-mean-square
values~\cite{PPlane}.  Another method, $v_2\{4\}$~\cite{v24} is supposed to better represent the
measurement with respect to the reaction plane.  For $20-50\%$ Au+Au collisions at 200 GeV, the
slope parameter obtained with $v_2\{4\}$ is illustrated with the
cross-hatched band in Fig.~\ref{fig:200GeV}, which is systematically lower
than the $v_2\{2\}$ results, but still has a finite positive value with a
larger statistical error.

Since the prediction of the consequence of CMW on $v_2$~\cite{CMW, CMW2}, this subject has
recently drawn increasing attention from
theorists~\cite{Yee2013Apr,Hirano,Wiedeman,Yee2013Nov,CMWnewSlope,Bloczynski2013,ChargeConserve}. It
was pointed out in Ref.~\cite{ChargeConserve} that local charge conservation at freeze-out, when
convoluted with the characteristic shape of $v_2(\pt)$ and $v_2(\eta)$, may provide a
qualitative explanation for the finite $v_2$ slope we observe. Such an effect depends on the
strength of the $\ach$ dependence on the mean $\pt$ and the $\eta$-dependence of $v_2$. However,
our measurements were carried out in a narrow $\pt$ range ([0.15,0.5] GeV/c) and with a
$\langle \pt \rangle(\ach)$ variation of 0.1\% at most. Furthermore, the measured
$\eta$-dependence of $v_2$ is only half as strong as that used in Ref.~\cite{ChargeConserve}.
We estimate the contribution of this mechanism to be smaller than the measurement by an order of
magnitude.

To check if the observed slope parameters come from conventional physics, such as Coulomb
interactions, or from a bias due to the analysis approach, we carried out the same analysis in
Monte Carlo events from UrQMD.  As shown in Fig.~\ref{fig:200GeV}, the slopes extracted from
UrQMD events of 200 GeV Au+Au collisions are consistent with zero for 10-60\% centrality
collisions, where the signal is prominent in the data. Similarly, the AMPT event
generator~\cite{ampt, Ma2014} also produces events with slopes $r$ consistent with zero. With
the AMPT model, we also studied the weak decay contribution to the slope, which was
negligible. On the other hand, the CMW calculations~\cite{CMW} demonstrate a similar centrality
dependence of the slope parameter.
Recently, a more realistic implementation of the CMW~\cite{Yee2013Nov} suggested that the CMW
contribution to $r$ is sizable, and the centrality dependence of $r$ is similar to
the data. In these theoretical calculations such centrality dependence mainly
  results from the centrality dependence of the magnetic field and the system volume.
Quantitative comparisons between data and theory require further work on both sides to match the
kinematic regions used in the analyses. For example, the measured $\ach$ only represents the
charge asymmetry of a slice ($|\eta|<1$) of an event, instead of that of the whole collision
system.  We expect these two values of $\ach$ to be proportional to each other, but the
determination of the ratio will be model dependent. In addition to the UrQMD and AMPT simulation
studies which reveal no trivial correlation between $\ach$ and pion $v_2$, tests were performed
using the experimental data. For example, $\ach$ and the pion $v_2$ were calculated in two
kinematically separated regions, i.e., different rapidity bins. In such cases, the slope
parameters decrease but remain significant and positive. This may reflect the local nature of
the $\ach$ dependence of $v_2$, but additional theoretical development is necessary.

Figure~\ref{fig:BES} shows a similar trend in the centrality dependence of the slope parameter
for all the beam energies except 11.5 and \unit{7.7}{GeV}, where the slopes are consistent with
zero with large statistical uncertainties. It was argued~\cite{quarkTransport} that at lower beam
energies the $\ach$-integrated $v_2$ difference between particles and anti-particles can be
explained by the effect of quark transport from the projectile nucleons to mid-rapidity,
assuming that the $v_2$ of transported quarks is larger than that of produced ones.  The same model,
however, when used to study $v_2(\pi^-) - v_2(\pi^+)$ as a function of $\ach$, suggested a
negative slope~\cite{quarkTransportQM12}, which is contradicted by the data.

The mean field potentials from the hadronic phase~\cite{hadronicPotential} and the partonic
phase~\cite{partonicMeanField} also qualitatively explain the $\ach$-integrated $v_2$
difference between particles and anti-particles, especially at lower beam energies.  In general,
the mean field potential is expected to be positively correlated with $\ach$ and thus may
explain the trends in those data, but no conclusive statement can be made here due to the lack
of specific predictions. This effect may be tested in the future by studying the $K^\pm$
$v_2$ slopes, whose $v_2$ ordering is opposite to that of $\pi^\pm$.

In summary, pion $v_2$ exhibits a linear dependence on $\ach$, with positive (negative) slopes
for $\pi^-$ ($\pi^+$). The $v_2(\pi^-) - v_2(\pi^+)$ increases as a function of $\ach$,
qualitatively reproducing the expectation from the CMW model. The slope $r$ of $v_2(\ach)$
difference between $\pi^-$ and $\pi^+$ has been studied as a function of centrality, and we
observe a dependence also similar to the calculation based on the CMW model.  The slope
parameter $r$ remains significantly positive for $10-60\%$ centrality Au+Au collisions at
$\sNN = 27-200$ GeV, and displays no obvious trend of the beam energy dependence with the
current statistics.  None of the conventional models discussed, as currently implemented, can
explain our observations.

We thank the RHIC Operations Group and RCF at BNL, the NERSC Center at LBNL, the KISTI Center in
Korea, and the Open Science Grid consortium for providing resources and support. This work was
supported in part by the Office of Nuclear Physics within the U.S. DOE Office of Science, the
U.S. NSF, the Ministry of Education and Science of the Russian Federation, NNSFC, the MoST of
China (973 Program No.~2014CB845400), CAS, MoST and MoE of China, the Korean Research
Foundation, GA and MSMT of the Czech Republic, FIAS of Germany, DAE, DST, and UGC of India, the
National Science Centre of Poland, National Research Foundation, the Ministry of Science,
Education and Sports of the Republic of Croatia, and RosAtom of Russia.

  
\end{document}

%% file: star_authors.tex
\affiliation{AGH University of Science and Technology, Cracow 30-059, Poland}
\affiliation{Argonne National Laboratory, Argonne, Illinois 60439, USA}
\affiliation{Brookhaven National Laboratory, Upton, New York 11973, USA}
\affiliation{University of California, Berkeley, California 94720, USA}
\affiliation{University of California, Davis, California 95616, USA}
\affiliation{University of California, Los Angeles, California 90095, USA}
\affiliation{Central China Normal University (HZNU), Wuhan 430079, China}
\affiliation{University of Illinois at Chicago, Chicago, Illinois 60607, USA}
\affiliation{Creighton University, Omaha, Nebraska 68178, USA}
\affiliation{Czech Technical University in Prague, FNSPE, Prague, 115 19, Czech Republic}
\affiliation{Nuclear Physics Institute AS CR, 250 68 \v{R}e\v{z}/Prague, Czech Republic}
\affiliation{Frankfurt Institute for Advanced Studies FIAS, Frankfurt 60438, Germany}
\affiliation{Institute of Physics, Bhubaneswar 751005, India}
\affiliation{Indian Institute of Technology, Mumbai 400076, India}
\affiliation{Indiana University, Bloomington, Indiana 47408, USA}
\affiliation{Alikhanov Institute for Theoretical and Experimental Physics, Moscow 117218, Russia}
\affiliation{University of Jammu, Jammu 180001, India}
\affiliation{Joint Institute for Nuclear Research, Dubna, 141 980, Russia}
\affiliation{Kent State University, Kent, Ohio 44242, USA}
\affiliation{University of Kentucky, Lexington, Kentucky, 40506-0055, USA}
\affiliation{Korea Institute of Science and Technology Information, Daejeon 305-701, Korea}
\affiliation{Institute of Modern Physics, Lanzhou 730000, China}
\affiliation{Lawrence Berkeley National Laboratory, Berkeley, California 94720, USA}
\affiliation{Max-Planck-Institut fur Physik, Munich 80805, Germany}
\affiliation{Michigan State University, East Lansing, Michigan 48824, USA}
\affiliation{Moscow Engineering Physics Institute, Moscow 115409, Russia}
\affiliation{National Institute of Science Education and Research, Bhubaneswar 751005, India}
\affiliation{Ohio State University, Columbus, Ohio 43210, USA}
\affiliation{Institute of Nuclear Physics PAN, Cracow 31-342, Poland}
\affiliation{Panjab University, Chandigarh 160014, India}
\affiliation{Pennsylvania State University, University Park, Pennsylvania 16802, USA}
\affiliation{Institute of High Energy Physics, Protvino 142281, Russia}
\affiliation{Purdue University, West Lafayette, Indiana 47907, USA}
\affiliation{Pusan National University, Pusan 609735, Republic of Korea}
\affiliation{University of Rajasthan, Jaipur 302004, India}
\affiliation{Rice University, Houston, Texas 77251, USA}
\affiliation{University of Science and Technology of China, Hefei 230026, China}
\affiliation{Shandong University, Jinan, Shandong 250100, China}
\affiliation{Shanghai Institute of Applied Physics, Shanghai 201800, China}
\affiliation{Temple University, Philadelphia, Pennsylvania 19122, USA}
\affiliation{Texas A\&M University, College Station, Texas 77843, USA}
\affiliation{University of Texas, Austin, Texas 78712, USA}
\affiliation{University of Houston, Houston, Texas 77204, USA}
\affiliation{Tsinghua University, Beijing 100084, China}
\affiliation{United States Naval Academy, Annapolis, Maryland, 21402, USA}
\affiliation{Valparaiso University, Valparaiso, Indiana 46383, USA}
\affiliation{Variable Energy Cyclotron Centre, Kolkata 700064, India}
\affiliation{Warsaw University of Technology, Warsaw 00-661, Poland}
\affiliation{Wayne State University, Detroit, Michigan 48201, USA}
\affiliation{World Laboratory for Cosmology and Particle Physics (WLCAPP), Cairo 11571, Egypt}
\affiliation{Yale University, New Haven, Connecticut 06520, USA}
\affiliation{University of Zagreb, Zagreb, HR-10002, Croatia}

\author{L.~Adamczyk}\affiliation{AGH University of Science and Technology, Cracow 30-059, Poland}
\author{J.~K.~Adkins}\affiliation{University of Kentucky, Lexington, Kentucky, 40506-0055, USA}
\author{G.~Agakishiev}\affiliation{Joint Institute for Nuclear Research, Dubna, 141 980, Russia}
\author{M.~M.~Aggarwal}\affiliation{Panjab University, Chandigarh 160014, India}
\author{Z.~Ahammed}\affiliation{Variable Energy Cyclotron Centre, Kolkata 700064, India}
\author{I.~Alekseev}\affiliation{Alikhanov Institute for Theoretical and Experimental Physics, Moscow 117218, Russia}
\author{J.~Alford}\affiliation{Kent State University, Kent, Ohio 44242, USA}
\author{A.~Aparin}\affiliation{Joint Institute for Nuclear Research, Dubna, 141 980, Russia}
\author{D.~Arkhipkin}\affiliation{Brookhaven National Laboratory, Upton, New York 11973, USA}
\author{E.~C.~Aschenauer}\affiliation{Brookhaven National Laboratory, Upton, New York 11973, USA}
% ...

\collaboration{STAR Collaboration}\noaffiliation